\documentclass[12pt]{article}
\usepackage{graphics}
\usepackage{epsfig}
\usepackage{amsfonts}
\usepackage{latexsym}
\usepackage{epic}
\newcommand{\ba}{\begin{array}}
\newcommand{\ea}{\end{array}}
\newcommand{\be}{\begin{equation}}
\newcommand{\ee}{\end{equation}}
\newcommand{\nn}{\nonumber}
\newcommand{\bea}{\begin{eqnarray}}
\newcommand{\eea}{\end{eqnarray}}
\newcommand{\beas}{\begin{eqnarray*}}
\newcommand{\enas}{\end{eqnarray*}}

\begin{document}
\begin{center}
{\Large 3D massive Dirac fermions with chemical potential in external magnetic field:
Current-current correlation function
}
\end{center}
\vspace{2.5 cm}
\begin{center}
{\bf  E. Apresyan\footnote{e-mail:{\sl
elena-apresyan@mail.ru}}} \\

\vspace{12pt}

{
\vspace{6pt}

Yerevan Physics Institute, Alikhanian Br. str. 2, Yerevan 36,  Armenia}
\end{center}
\vspace{1.5 cm}

\begin{center}
{\Large  Abstract}
\end{center}

 The response of fermionic system to external gauge fields in presence
 of non-quantized magnetic field is determined by current-current correlation
 function $\Pi_{\mu\nu}({\bf B})$. We study $2D$ dimensional Dirac electron system
  and calculate current-current
 correlation function in a presence of  magnetic field $B$,
 chemical potential $\eta$ and gap $m$.

\newpage
\section{Introduction}

The experimental and theoretical study of graphene,two-dimensional graphite,
is an extremely rapidly growing field of today's condensed matter research.
The reasons for enormous scientific interest are manifold.Graphene is a zero
gap semiconductor because its conduction and valence bands meet at the Dirac point
\cite{CastroNeto-1-2008}.Electronic properties of graphene is sensitive to environmental conditions therefore
there will be changed in presence of other layers.Graphene has peculiar band structure as a result electrons
at Fermi energy are discribed an effective Lorentz invariant theory.
Electrons propogating through graphen's honeycomb lattice effectively lose their
mass,in result producing quasi-particles that are described $2D$ analogue of Dirac equation \cite{Novoselov-2005}.
The theoretical and experimental studies of the influence of the external fields on
the graphene transport features are held recently \cite{Tahir, Peres-2006}.
The constant magnetic field acts as a strong catalyst  of dynamical symmetry breaking leadind to the generation
of fermion masses in $2+1$ dimension.There is astriking similarity between the role of  magnetic field in $2+1$ dimensional models and
the role of Fermi surface in the Bardeen-Cooper-Schrieffer (BCS) theory of superconductivity \cite{J.Barden, N.L.Cooper}.
The magnetic field influences on
the high frequency conductivity and on the electromagnetic waves absorption of the
graphene are investigated in.Magnetic field applied to graphene gives rise to descrete Landau levels
which are essentially important in the explanation of a anomalous quantum Hall effect in graphene
\cite{Laughlin-1981,Jakiw,Zigler}.By analyzing properties of quantum Hall effect in a weak field was obtained
that low energy exitations in graphene are Dirac quasiparticles.
The dependence of Hall conductivity  on the magnetic field intensity was investigated.
The graphene conductivity have the oscillations when the magnetic field intensity changes \cite{Kryuchkov-2012}
However,current response functions are not fully studied in a presence of magnetic field.
In the paper \cite{Khalilov} the transversal part of the polarization operator  $\Pi^{\mu \nu}$ was calculated.
The goal in  this work is the calculation of current density correlation function $\Pi^{\mu 0}$
when we have third order Feynman's diagram in presence of gap  $m$,$\eta$ chemical potential and $B$ magnetic field.
Exact expression of polarization operator without magnetic field but for finite chemical potential $\eta$ and gap $m$
was calculated in \cite{Apresyan}.


\section{Current-current correlation function $\Pi^{\mu\nu}({\bf B})$}

The action which describes the graphene in the Effective Field Theory (EFT)  framework
 via $N_f$ four-component massive Dirac fermions with instantaneous three-dimensional
 Coulomb interactions is the
 following (in Euclidean space time) \cite{Semenoff, Son}
 \bea
\label{action} S_g=-\sum_{i=1}^{N_f}\int {d^2 x  d t}\bar\psi_i
\left(\gamma^0 \partial_0+v \gamma^k\partial_k+i A_0\gamma^0 +m
\right)\psi_i+\frac{1}{2g^2}\int {d^2 x  d t}(\partial_k A_\mu)^2.
\eea
Here $v$ is the velocity, which can be taken as $1$ in the
calculations and then restore in the resulting formulas. In real
graphene $N_f=2$, $\gamma$-matrices satisfy to Euclidean Clifford
algebra and can be chosen as
\bea \gamma^0=\sigma^3\otimes \sigma^3,\;\;\;
\gamma^i=\sigma^i\otimes 1,\;\;\;\{\gamma^\mu
\gamma^\nu\}=2\delta^{\mu\nu}. \eea
The four-component fermionic structure is conditioned by the
existence of the quasi-particle excitations in two sublattices in
the graphene around two Dirac points.

Since each Dirac point contributes to response function
additively,  below, for simplicity, we will be concentrated on
calculation of current-current correlation function only  for
single Dirac point. Therefore we start from {\it{free}} Dirac
action in three dimensional space-time with chemical potential
$\eta$ ,gap $m$ and $B$ magnetic field, which after Wick rotation to complex
time/energy acquires the form
\bea \label{D} S=\int \frac{d {\bf
k} d \omega}{(2\pi)^3} \bar\psi_{\bf k, \omega}[{\bf \sigma} {\bf
k}+ \sigma_3  m -(\omega -i \eta)]\psi_{{\bf k}, \omega}, \eea

where the Fourier transformation is done (${\bf{k}}=\{k_1,k_2\}$)
and in the role of $\gamma$ functions Pauli matrices are taken.
Here we intend to calculate the current-current correlation function for the
three-dimensional theory with the kinetic part for the
fermions presented above and the interaction term with
$U(1)$ gauge field $A_\mu$ in the third order  approximation. 

\begin{figure}[t]
\centerline
{\includegraphics[width=90mm,angle=0,clip ]{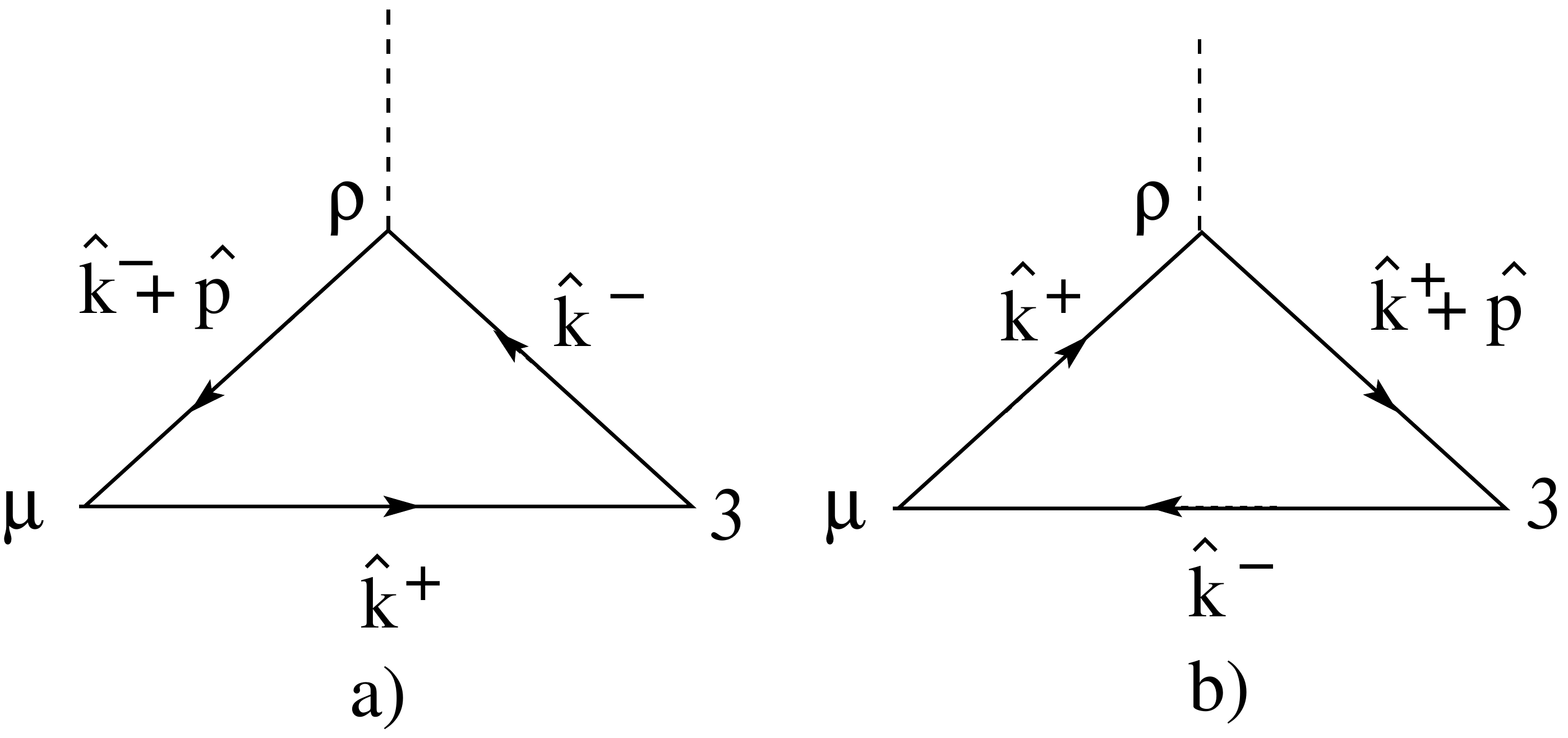}}
\caption{Third order Feynman diagram for current-current correlation function . }
\label{fig-diagram}
\end{figure}

The magnetic field dependence of the current-current correlation function is defined by third order
Feynman diagrams in Fig.$\ref{fig-diagram}$,
where vector potential $A_\rho$ couples to vertex $\rho$. After some transformations  diagram a) reads

\bea
\label{Pi}
 \Pi_{\mu 0}&=&Ng^2\int_{-\infty}^{+\infty}\frac{d^3 k}{(2\pi)^3}{\mbox{Tr}}[\sigma_\mu G(\hat k^+ )A_\rho\sigma_\rho G(\hat
{k}^++\hat{p})\sigma_3 G(\hat{k}^{-})]\nn\\
 &=&N g^2\int_{-\infty}^{+\infty}\frac{d^3k}{(2\pi)^3}
 {\mbox{Tr}}[\sigma_\mu G(\hat {k}^+) A_{\rho}\sigma_\rho \frac{\hat{p}}{p^2+m^2} \sigma_3 G(\hat{k}^-)]
\eea
where $G(\hat k)=\frac{\hat k -m}{k^2+m^2}$ is the Green function of the fermion and we have used the notation $k^\pm=(\vec{k}\pm\frac{\vec{q}}{2},\Omega\pm\frac{\omega}{2})$. By using identity
$A_\rho \sigma_\rho \hat{p}=\vec{A}\vec{p}+i\epsilon_{\nu\rho}A_\nu p_\rho\sigma_3 = i B \sigma_3$ in second row of the expression (\ref{Pi}),
where we have dropped $\vec{A}\vec{p} $ term since it gives zero, we come to following Trace in the nominator
 \bea
\label{trace}
 &B \mbox{Tr}[\sigma_\mu (\hat{k}^+-m)\sigma_3 \sigma_3(\hat{k}^--m)]=
 2B(\epsilon_{\mu \nu \sigma} k_\nu^+ k_\sigma^- - m (k^+ + k^-)_\mu)\\ \nn
&=2B[\epsilon_{\mu\nu}(q_\nu \Omega-k_\nu \omega)-2 m k_\mu]
\eea

In the same way one can find corresponding expression for Trace for diagram of Fig.$1(b)$, which coincides with (\ref{trace}).

We see, that in three dimensional space the third order Feynman's diagrams are not vanish, therefore, summarizing Trace results we obtain
$4B[\epsilon_{\mu\nu}(q_\nu \Omega-k_\nu \omega)-2 m k_\mu]$.

\section{Calculation of $\Pi_{\mu 0}$}

Using the trace result  $\Pi_{\mu 0}$ is acquire following form
\bea
\label{ten}
\Pi_{\mu 0}(B)=Ng^2\int_{-\infty}^{+\infty}\frac{d^3k}{(2\pi)^3}
\Big(\frac{4B[\epsilon_{\mu\nu}(q_{\nu}\Omega-k_{\nu}\omega)-2mk_{\mu}]}
{(k^{+2}+m^2)(k^{-2}+m^2)^2}+\frac{4B[\epsilon_{\mu\nu}(q_{\nu}\Omega-k_{\nu}\omega)-2mk_{\mu}]}{(k^{-2}+m^2)(k^{+2}+m^2)^2}\Big)
\eea
where $k^2=\vec{k^2}+(\Omega+\Gamma+i\eta)$.
Generally $\Pi_{\mu\nu}$ must satisfy the condition of conservation
of charge  $\partial_\mu \Pi_{\mu \nu}=0$.
The evaluation of such integrals performs with the method of Feynman parametrization.
These method gives opportunity to squeeze the three  denominator factors into single quadratic as polynomial of $k$.
After we shift $k$ by a constant. It is easy to begin with trivial case when in denominator we have two factors
\bea
\label{six}
\frac{1}{AB}=2\int_{0}^{1}dx_1dx_2\frac{\delta(x_1+x_2-1)}{[x_1A+x_2B]^2}
\eea
When we have three factors then
\bea
\label{3}
\frac{1}{ABC}=2\int_{0}^{1}dx_1dx_2dx_3\frac{\delta(x_1+x_2+x_3-1)}{[x_1A+x_2B+x_3C]^3}
\eea
Our integral (\ref{ten}) has three factors in the denominator, therefore by using (\ref{3}) we obtain
\bea
\label{bill}
\frac{1}{AB^2}=\frac{\Gamma(1+2)}{\Gamma(1)\Gamma(2)}
\int_{0}^{1}du_1du_2\frac{\delta(u_1+u_2-1)u_2}{(u_1A+u_2B)^3}=2!\int_{0}^{1}
du\frac{1-u}{(uA+(1-u)B)^3}
\eea
where $A=k^{-2}+m^2$, $B=k^{+2}+m^2$. Easy to find out, that making shift
$k^{\pm} = k^{'\pm} +(1/2 - u) q$ we come to very simple expressions

\bea\label{one}
\frac{1}{AB^2}&=&2!\int_{0}^{1}du \frac{1-u}{[k^{'2}+u(1-u)q^2+m^2]^3}\nn\\
\frac{1}{A^2B}&=&2!\int_{0}^{1}du \frac{u}{[k^{'2}+u(1-u)q^2+m^2]^3}.
\eea
Then the polarization operator $\Pi_{\mu 0}$ defined by (\ref{ten}) acquires the form
\bea
\label{five}
&&\Pi_{\mu 0}(B)\nn \\
&=& 8B\int_{0}^{1}du\frac{d^3k^{'}}{(2\pi)^3}\frac{[\epsilon_{\mu\nu}q_{\nu}(\Omega^{'}+(\frac{1}{2}-u)\omega)-
(k^{'}+(\frac{1}{2}-u)q)_{\nu}\omega-2m(k^{'}+(\frac{1}{2}-u)q)_{\nu}]}{(k^{'2}+m^2+u(1-u)q^2)^3}\nn \\
&=&8B\int_{0}^{1}du\frac{d^3k}{(2\pi)^3}\frac{\epsilon_{\mu\nu}
[q_\nu(\frac{1}{2}-u)(\Gamma+i\eta)-(\frac{1}{2}-u)q_{\nu\omega}]-2m(\frac{1}{2}-u)q_{\mu}}{(k^2+m^2+u(1-u)q^2)^3}\\ \nn
&=&8B\int_{0}^{1}du\frac{d^2k}{(2\pi)^2}\frac{d \Omega}{2\pi}
\frac{\epsilon_{\mu\nu}q_{\nu}(\Gamma+i\eta)-m(1-2u)q_{\mu}}{\Big[(\Omega+\Gamma+i\eta)-\sqrt{\vec{k}^2+m^2+u(1-u)q^2}\Big]^3}\\ \nn
&&\times\frac{1}{\Big[(\Omega+\Gamma+i\eta)+\sqrt{\vec{k}^2+m^2+u(1-u)q^2}\Big]^3}\nn
\eea

In (\ref{five}) we see, that have a pole of third order, therefore, applying  Cauchy integration formula
and differentiating twice integrand of (\ref{five}) over $\Omega$ we obtain
\bea
\label{div}
\Pi_{\mu 0}(B)&=&8 i B \int_{0}^{1}du\frac{d\vec{k}}{(2\pi)^2}\frac{\partial^2}{\partial\Omega^2}
\frac{\epsilon_{\mu\nu}q_{\nu}(\Gamma+i\eta)-m(1-2u)q_{\mu}}{[(\Omega+\Gamma+i\eta)-\sqrt{\vec{k}^2+m^2+u(1-u)q^2}]^3}\\ \nn
&&\times\frac{1}{[(\Omega+\Gamma+i\eta)+\sqrt{\vec{k}^2+m^2+u(1-u)q^2}]^3}\\ \nn
&=& \frac{3 i B}{2}\int_{0}^{1}du\frac{d\vec{k}}{(2\pi)^2}
\frac{\epsilon_{\mu\nu}q_{\nu}(\Gamma+i\eta)-m(1-2u)q_{\mu}}{(\vec{k}+m^2+u(1-u)q^2)^{5/2}}
\eea
Now, by performing integration over $\vec{k}$ using standard formula of dimensional regularization
\bea
\int\frac{d^2k}{(2\pi)^d}\frac{1}{(k^2+\Delta)^n}=\frac{1}{(4\pi)^{\frac{d}{2}}}\frac{\Gamma(n-\frac{d}{2})}{\Gamma(n)}\frac{1}{\Delta^{n-\frac{d}{2}}}
\eea
 and dividing the range of integration $[0,1]$ in three part  we obtain
\bea
&&\Pi_{\mu 0}(B) =\nn\\
&=&\frac{i B}{4\pi}\Bigg\{\int_{u_1}^{u_2}du\frac{\epsilon_{\mu\nu}q_{\nu}(\Gamma+i\eta)-m(1-2u)q_{\mu}}{(m^2+u(1-u)q^2)^{3/2}}
-\int_{0}^{u^1}du\frac{\epsilon_{\mu\nu}q_{\nu}(\Gamma+i\eta)-m(1-2u)q_{\mu}}{\eta^3}-\\ \nn
&&\int_{u_2}^{1}du\frac{\epsilon_{\mu\nu}q_{\nu}(\Gamma+i\eta)-m(1-2u)q_{\mu}}{\eta^3}\Bigg\}
=-\frac{i B}{4\pi}\Bigg[\Bigg(-2\frac{\frac{1-2u}{4m^2+q^2}\epsilon_{\mu\nu}q_{\nu}
(\Gamma+i\eta)+2m\frac{q_\mu}{q^2}}{(m^2+u(1-u)q^2)^{1/2}}|_{u_1}^{u_2}\Bigg)\\ \nn
&+&\frac{1}{\eta^3}\epsilon_{\mu\nu}q_\mu(\Gamma+i\eta)(1+u_1-u_2)+\frac{m}{\eta^3}q_{\mu}(u_1-u_2)(u_1+u_2-1)\Bigg]\\ \nn
&=&-\frac{Bi}{\pi}\frac{\epsilon_{\mu\nu}q_{\nu}(\Gamma+i\eta)}{(4m^2+q^2)|\eta|}\sqrt{1-\frac{4(\eta^2-m^2)}{q^2}}-
\frac{Bi}{4\pi|\eta|}\frac{\epsilon_{\mu\nu}q_{\nu}}{\eta^3}(\Gamma+i\eta)\Bigg[1-\sqrt{1-\frac{4(\eta^2-m^2)}{q^2}}\Bigg]\\ \nn
\label{ccc}
\eea
where expressions $u_1=\frac{1}{2}\Big(1-\sqrt{1-\frac{4(\eta^2-m^2)}{q^2}}\Big)$, $u_2=\frac{1}{2}\Big(1+\sqrt{1-\frac{4(\eta^2-m^2)}{q^2}}\Big)$
are obtained from the equation $m^2+u(1-u)q^2=\eta^2$.

\section{Results}

Finally, in case of $\frac{q^2}{4}\geq(\eta^2-m^2)\geq 0 $, when the square root in the expression of $u_{1,2}$ is real,
the integral over $u$ gives
\bea \label{three}
\Pi_{\mu 0}(B)=-\frac{iB}{4\pi | \eta |}\epsilon_{\mu\nu}q_{\nu}(\Gamma+i\eta)
\Big(\frac{1}{m^2+\frac{q^2}{4}} \sqrt{1-\frac{4(\eta^2-m^2)}{q^2}}+ \frac{1}{\eta^2}\Big(1-\sqrt{1-\frac{4(\eta^2-m^2)}{q^2}}\Big)\Big)
\eea
Denote that for  polarization operator take place the condition of conservation of charge.
 When $\eta^2-m^2 \geq \frac{q^2}{4}$, then  $u_1=u_2=\frac{1}{2}$ and  for $\Pi_{\mu 3}(B)$ we obtain
\bea
\Pi_{\mu 0}(B)=-\frac{iB}{4\pi|\eta|^3}\epsilon_{\mu\nu}q_{\nu}(\Gamma+i\eta)
\eea
For $\eta^2-m^2\leq 0$ then $u_1=0, u_2=1$ and in a result we have following expression
\bea
\Pi_{\mu\nu}(B)=-\frac{iB}{\pi m}\epsilon_{\mu\nu}q_{\nu}(\Gamma+i\eta)\frac{1}{4m^2+q^2}
\eea

\section{Acknowledgment}
I would like to express my sincere gratitude to my advisor Prof.A.Sedrakyan for the continuous support,
for his patience, motivation, and immense knowledge. His guidance helped me for writing this article.
The work was supported by ARC grant 15T-1C058.


\begin{thebibliography}{99}

\bibitem{CastroNeto-1-2008}  J. Sabio, J. Nilsson, and A. H. Castro Neto, Phys. Rev. B 78, 075410 (2008).

\bibitem{Novoselov-2005} K. Novoselov,  A. Geim-Nature, 2005.

\bibitem{Tahir} M.Tahir and K.Sabeeh Phys. Rev. B, Vol. 77, No.19.

\bibitem{Peres-2006} N.M.R.Peres,F. Guinea and A.H.C. Neto, Phys. Rev.B Vol. 73, 2006 .

\bibitem{J.Barden} J.Barden,N.L.Cooper,J.R.Schrieffer, Theory of superconductivity Phys.Rev.(108)(1957) 1175-1204 doi:10.1103/PhysRev.108.1175

\bibitem{N.L.Cooper}J. R. Schrieffer, Microscopic theory of superconductivity, Phys. Rev. 106 (1957) 162.

doi:10.1103/PhysRev.106.162.

\bibitem {Laughlin-1981}Laughlin, R. (1981). "Quantized Hall conductivity in two dimensions". Physical Review B

\bibitem{Jakiw} R.Jakiw-Phys. Rev. {\bf D 29 }, 2375 (1984) .

\bibitem{Zigler} K. Zigler, Phys. Rev. Lett. 80, 3113 (1998).

\bibitem{Kryuchkov-2012} S.V.Kryuchkov,E.I.Kukhar Journal of Modern Physics,2012,3, 994-1001.

\bibitem{Khalilov} V.R. Khalilov and I.V. Mamsurov, Eur. Phys. J. C 75, 167 (2015)

\bibitem{Apresyan} E. Apresyan , Sh. Khachatryan and A. Sedrakyan- Mod. Phys. Lett. A 30, 1550035 (2015).

\bibitem{Semenoff} G. Semenoff, Phys. Rev. Lett. 53, 2449 (1984).

\bibitem{Son} D. T. Son, Phys. Rev. {\bf B 75}, 235423 (2007).

\end{thebibliography}
\end{document}